\documentclass[aps,prl,reprint,superscriptaddress]{revtex4-2}
\usepackage{braket}
\usepackage{quantikz}
\usepackage{graphicx}
\usepackage{dcolumn}
\usepackage{bm}
\usepackage{lipsum}
\usepackage{amsthm}
\usepackage{siunitx}
\usepackage{physics}
\usepackage{amssymb}
\usepackage{placeins}
\usepackage{comment}

\usepackage[normalem]{ulem} 
\usepackage[colorlinks,linkcolor=blue,urlcolor=blue, citecolor=blue]{hyperref}
\def\captionof#1#2{{\def\@captype{#1}#2}}

\begin{document}

\preprint{APS/123-QED}

\title{Bright Source of High-Dimensional Temporal Entanglement}

\author{Dorian Schiffer}
\email{Dorian.Schiffer@tuwien.ac.at}
\thanks{These authors contributed equally to this work.}
\affiliation{Atominstitut, Technische Universit\"at Wien, Stadionallee 2, 1020 Vienna, Austria}
\affiliation{Institute for Quantum Optics and Quantum Information (IQOQI), Austrian Academy of Sciences,
Boltzmanngasse 3, 1090 Vienna, Austria}

\author{Robert Kindler}
\email{Robert.Kindler@oeaw.ac.at}
\thanks{These authors contributed equally to this work.}
\affiliation{Institute for Quantum Optics and Quantum Information (IQOQI), Austrian Academy of Sciences,
Boltzmanngasse 3, 1090 Vienna, Austria}

\author{Alexandra Bergmayr-Mann}
\affiliation{Atominstitut, Technische Universit\"at Wien, Stadionallee 2, 1020 Vienna, Austria}

\author{Florian Kanitschar}
\affiliation{Atominstitut, Technische Universit\"at Wien, Stadionallee 2, 1020 Vienna, Austria}
\affiliation{AIT Austrian Institute of Technology, Center for Digital Safety $\&$ Security, Giefinggasse 4, 1210 Vienna, Austria}

\author{Amin Babazadeh}
\affiliation{Institute for Quantum Optics and Quantum Information (IQOQI), Austrian Academy of Sciences,
Boltzmanngasse 3, 1090 Vienna, Austria}

\author{Paul Erker}
\email{Paul.Erker@tuwien.ac.at}
\affiliation{Atominstitut, Technische Universit\"at Wien, Stadionallee 2, 1020 Vienna, Austria}
\affiliation{Institute for Quantum Optics and Quantum Information (IQOQI), Austrian Academy of Sciences,
Boltzmanngasse 3, 1090 Vienna, Austria}

\author{Marcus Huber}
\email{Marcus.Huber@tuwien.ac.at}
\affiliation{Atominstitut, Technische Universit\"at Wien, Stadionallee 2, 1020 Vienna, Austria}
\affiliation{Institute for Quantum Optics and Quantum Information (IQOQI), Austrian Academy of Sciences,
Boltzmanngasse 3, 1090 Vienna, Austria}

\author{Anton Zeilinger}
\email{Anton.Zeilinger@oeaw.ac.at}
\affiliation{Institute for Quantum Optics and Quantum Information (IQOQI), Austrian Academy of Sciences,
Boltzmanngasse 3, 1090 Vienna, Austria}

\date{\today}

\begin{abstract}
High-dimensional entanglement is considered to hold great potential for quantum key distribution (QKD) in high-loss and -noise scenarios. To harness its robustness, we construct a source for high-dimensional time-bin entangled photons optimized for high brightness, low complexity, and long-term stability. We certify the generated high-dimensional entanglement with a new witness employing nested Franson interferometry. Finally, we obtain key rates using a novel, noise-resilient QKD protocol. Our flexible evaluation method, centered around discretizations of the time stream, enables the same dataset to be processed while varying parameters such as state dimensionality and time bin length, allowing optimization of performance under given environmental conditions. Our results indicate regions within the accessible parameter space where high key rates per time are achievable for dimensionalities larger than two.
\end{abstract}

\maketitle

\section{I. Introduction}
Entanglement stands at the center of quantum physics. Since its discovery, this puzzling phenomenon has inspired physicists and philosophers alike to probe the foundations of physics \cite{Einstein1935, Bell1964, Freedman1972,Bell:1980wg, Aspect1981, Aspect1982a, Aspect1982b,Bouwmeester1997, Weihs1998, Bertlmann_2014, Giustina2015}. With the maturing of quantum technologies, however, entanglement is more and more regarded as a resource. As with any other physical resource, once produced, the problem of distribution arises. For quantum resources, which are notoriously fragile, solving this problem is a crucial step towards the quantum technology roll-out. In the context of Quantum Key Distribution (QKD) \cite{Bennett_Brassard1984, Ekert1991, Pirandola2020}, the distribution of the quantum resource, usually photons, in a lossy and noisy environment remains the main bottleneck for successful protocol implementation. Especially for entanglement-based QKD this problem stands at the center of current research \cite{review1}. Recently, high-dimensional (HD) entanglement emerged as a promising resource for QKD given its inherent resilience to noise \cite{zhu2020}. Several photonic degrees of freedom (DOFs) can host HD entanglement, including spatial structure, path, and time-energy \cite{mair2001, Fickler2012,fickler2014,bavaresco2020,Hu_2020,kaichi1,Anton1,Cabrejo_Ponce_2023,Serino2024}. The latter is of particular interest to us, since temporal entanglement, and its discretized version \emph{time-bin entanglement}, survive free-space and fiber channels and exhibit great adaptability to different loss and noise scenarios \cite{Brendel1999, ecker2019}.

Historically, most entanglement-based QKD protocols relied on a maximally polarization-entangled Bell state \cite{Gisin2002}. However, this approach has several drawbacks: For one, the dimensionality of polarization entanglement is strictly limited to that of qubits. Furthermore, high-brightness sources for polarization entanglement typically require interferometric stability, e.g., in a Sagnac configuration \cite{Kim2006, Hentschel2009}, thus severely reducing the long-term stability. Such sources face serious disadvantages in real-life applications, for instance, when deployed on a satellite for long-distance QKD. Protocols relying on hyper entanglement in the polarization and temporal DOFs represent a step forward. Employing such a strategy, successful key distribution over a $\SI{10.2}{\kilo \meter}$ free-space channel at dawn was demonstrated~\cite{bulla2022, bulla2023}. However, the requirement of these approaches for polarization entanglement implies that they inherit the same limitations described above.

In contrast, our QKD protocol, reported in Ref.~\cite{bergmayr2023}, is agnostic regarding the polarization DOF, and hence avoids these limitations. 
Using only time-bin entanglement allows us to construct a low-complexity entangled-photon source that can be simultaneously optimized for high brightness and good heralding efficiency with significantly fewer technical constraints. The low complexity enables source stability over extended periods of time without realignment. The main challenge, however, lies in demonstrating the entanglement quality and dimensionality of the generated state. Consequently, describing the source independently of the entanglement certification and benchmarking setup is of limited value.

To characterize our source, we employ our QKD protocol specifically adapted to the experiment and exploit a novel entanglement witness. The detection setup is therefore at least as crucial as the source itself. We are implementing the required measurements using a \textit{nested Franson interferomenter} with different delays. In our protocol, we record only the time-of-arrival statistics from all detectors across all measurement settings. This allows for the optimization of different figures of merit in post-processing by exploring various discretizations of the same raw data. In principle, this flexibility enables real-time adaptation to varying noise and loss conditions, paving the way toward full daytime QKD over free-space channels, such as, e.g., satellite-based quantum communications links \cite{liao2018,satelite1,satelite2}.

\begin{figure*}[htp!]
\begin{center}
\includegraphics[width=0.98\textwidth]{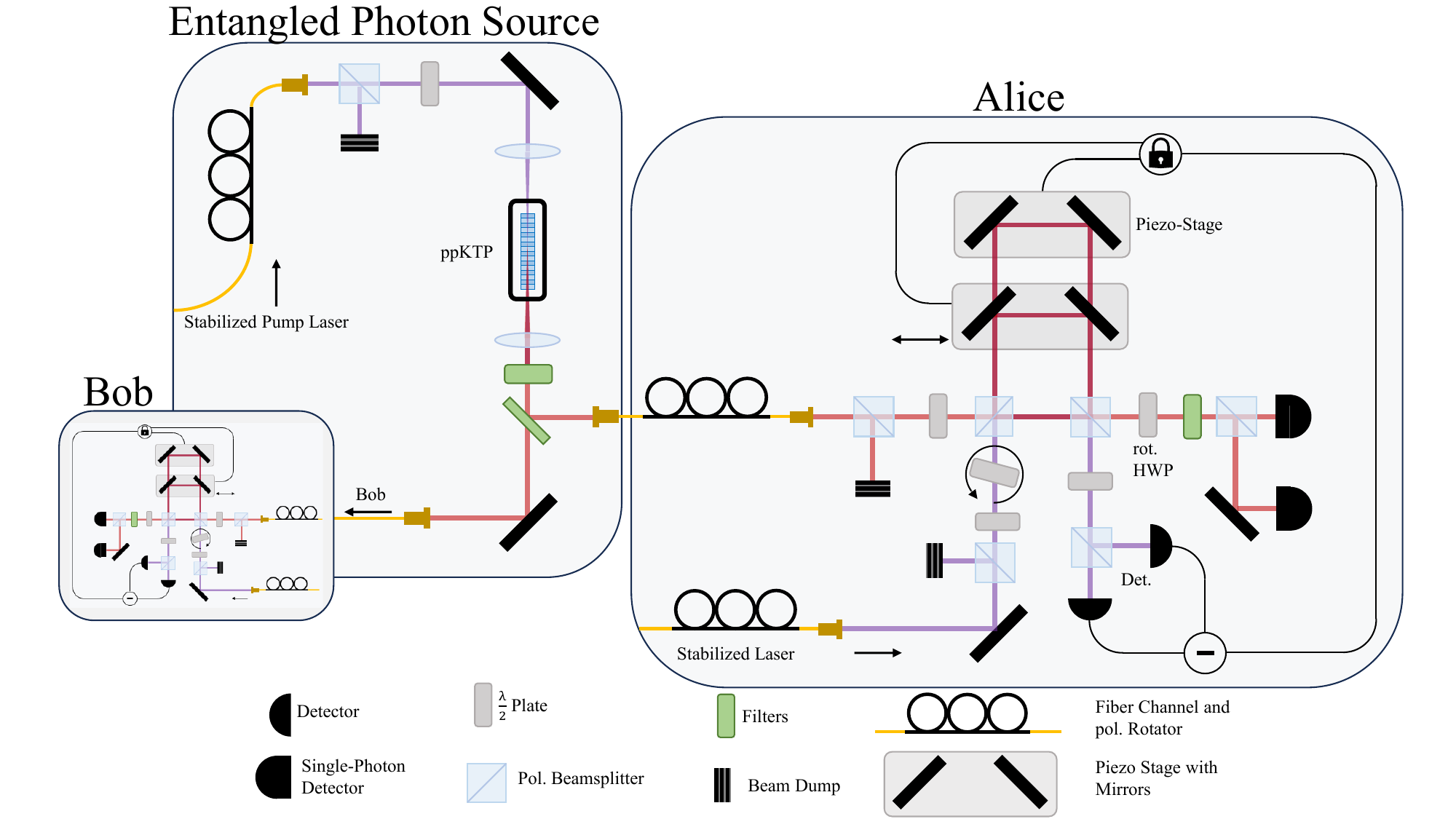}
\end{center}
\caption{Experimental Setup. In the photon source, we pump a Type-0 ppKTP crystal, kept at a non-degenerate temperature, with a wavelength-locked \SI{404.5}{nm} laser. After focusing the pump into the crystal, we collimate the generated photons and filter out the pump before splitting the photon pairs apart using a dichroic mirror. We couple the photons into single-mode fibers and distribute them to the detection modules, Alice and Bob. The PBSs and HWPs in front of the MZIs impose diagonal polarization and act as additional filters of potential background noise. Each module consists of two nested, imbalanced MZIs constructed using PBSs. By rotating a HWP in the output of the MZIs, we switch between time-of-arrival and time-superposition measurements. Alternatively, the HWPs in front of the MZIs could be rotated to achieve an equivalent effect. By monitoring the classical interference of the stabilized reference (pump) laser sent through the MZIs, we stabilize the interferometers using a piezo-actuated mirror stage.}
\label{fig:setup}
\end{figure*}

\section{II. Experimental Setup}
In this section, we first describe and characterize our new HD time-bin entangled-photon source and continue by detailing the nested-Franson detection apparatus used to assess the entanglement properties -- specifically the Schmidt number and entanglement rate -- of the generated photonic states and to demonstrate a proof-of-principle realization of the novel QKD protocol. The experimental setup is depicted in Figure \ref{fig:setup}.

\subsection{A. Entangled-Photon Source}
We generate HD entangled photons via spontaneous parametric down-conversion (SPDC) in a periodically poled Potassium Titanyl Phosphate (ppKTP) crystal. For operation under severe loss and high noise conditions, the novel source must satisfy several design requirements.

In order to achieve \emph{high brightness}, we employ a \SI{3}{cm} Type-0 ppKTP crystal, whose significantly larger non-linear coefficient gives us an advantage over Type-II crystals \cite{steinlechner2014}. However, by doing so, we cannot deterministically separate signal and idler photons by polarization when running the source at degenerate temperature. Therefore, we tune the crystal temperature to shift the wavelengths of the down-converted photons to allow separation by a dichroic mirror.

To achieve \emph{optimal heralding efficiency} without significantly suppressing brightness, we choose focal parameters according to \cite{bennink2010}, which determine the focal lengths of the lenses focusing the pump beam into the crystal and coupling out the photon pairs. Since pump power is not a limiting factor in our setup, we chose relatively low focal parameters ($\approx 0.2$). Lower values would further reduce the brightness while requiring even longer focal lengths and resulting in a prohibitively large physical footprint of the source.

We pump the crystal with a  $\SI{404.53}{\nano \meter }$ $cw$-laser locked to a hyperfine-transition in Potassium, yielding a wavelength stability of around $\SI{3}{\femto \meter}$ per hour over 24 hours. The SPDC spectrum is centered at $\lambda_s = \SI{774.6}{\nano \meter}$ for the signal photons and at $\lambda_i = \SI{845.4} {\nano \meter}$ for the idler photons with line-widths of $\SI{2.2}{\nano \meter}$ and $\SI{2.3}{\nano \meter}$ respectively.

We reach count rates of $2.2\times10^6$ coincidences/s/mW and measure a symmetric heralding efficiency of $\approx 30\%$ at $\SI{0.1}{\milli \watt}$ pump power. Note that without spectral filtering, there is an inherent trade-off between coincidence rate and heralding efficiency. our heralding efficiency is mainly limited by the relatively low detection efficiency ($\approx 60\%$) of our avalanche photodiodes (APDs).

The brightness of our source is not constrained by available pump laser power of up to $\SI{50}{\milli \watt}$, but rather by the detector dead-time, which causes a non-linear detector response already at low pump powers corresponding to a few million counts per second. Therefore, the full potential of our source can only be harnessed in links with very high transmission losses or, in principle, under tight spectral filtering.

\subsection{B. Detection Apparatus}

To verify that the photons emitted by our source are in fact HD entangled, we employ a polarizing nested-Franson-type setup. Conveniently, the same apparatus enables the implementation of the novel QKD protocol.

The photons travel from the source to each detection module through SM fibers and pass a manual polarization controller followed by a free-space filter stage that sets the polarization state for both photons to $\ket{D}$. The centerpiece of each detection module is two nested, imbalanced Mach-Zehnder interferometers (MZIs) followed by time-resolving photon detection. By discretizing the time stream into \emph{time bins} of length $\tau$, we detect HD entanglement in the time domain and gauge key rates.

After the MZIs, photons enter the analysis stage, where a rotatable half-wave plate (HWP) selects whether they are measured in the H/V or the D/A basis. In the former case, we obtain which-way information about the MZI and deterministically assign each click event a time bin. This implements a measurement in the \emph{Time-Of-Arrival} (TOA) basis. Selecting the D/A basis instead projects on superpositions of different time bins and realized measurements in the \emph{Time Superposition} (TSUP) basis.

Our nested Franson interferometer \cite{franson1989} introduces a temporal delay of $\approx \SI{1.3}{\nano \second}$ for the shorter and $\approx \SI{2.6}{\nano \second}$ for the longer arm. This allows discretizations of the time stream where the short MZI overlaps time bins $\ket{i}$ and $\ket{i-1}$ and the long MZI time bins $\ket{i}$ and $\ket{i-2}$, thereby accessing more superposition basis elements.

\begin{figure}[htp!]
\begin{center}
\includegraphics[width=\columnwidth]{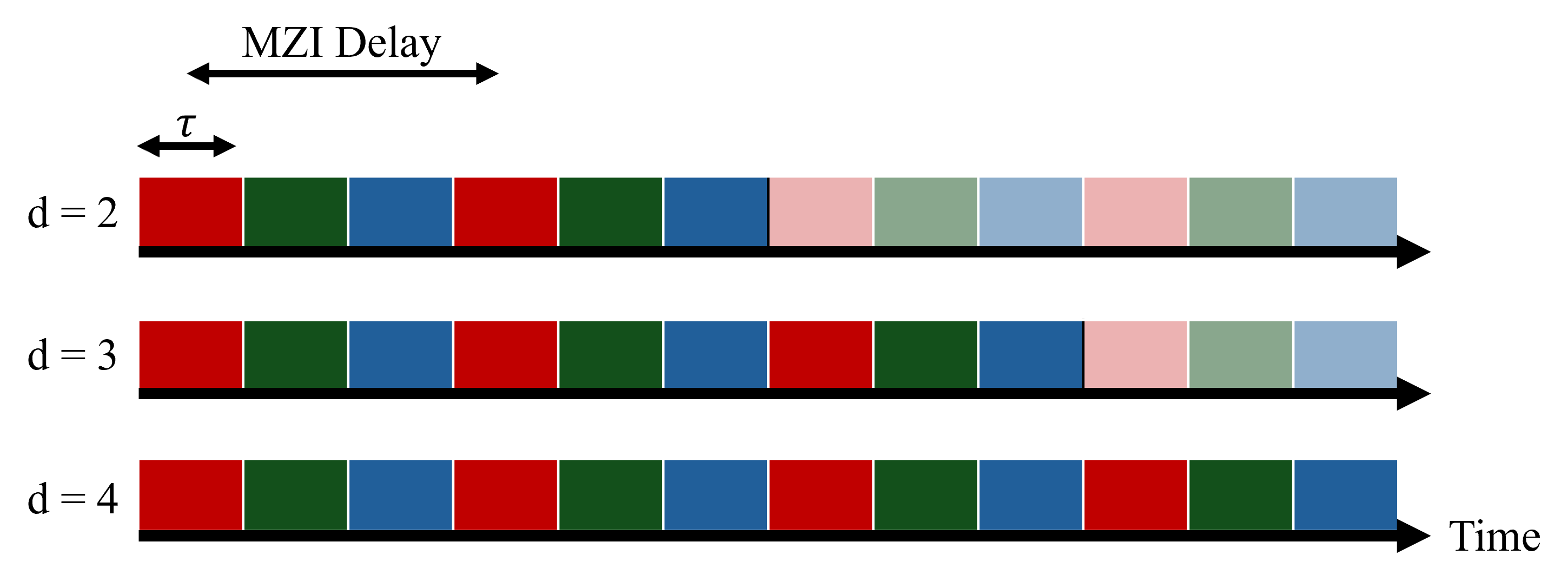}
\end{center}
\caption{Discretizations of the time stream for different dimensions. We group $d$ time bins of length $\tau$ into interleaved time frames, indicated by identical colors. Time bins belonging to the same state are separated by one MZI delay to enable TSUP measurements. This structure requires the MZI delay to be an integer multiple of $\tau$. After $d$ MZI delays, the pattern repeats and a new block of interleaved time frames begins, indicated in grayed-out colors.}
\label{fig:interleaved_TB}
\end{figure}

To prevent losses, our discretization strategy \cite{bulla2023} groups $d$ time bins in interleaved \emph{time frames}, thus covering the whole time stream, which constrains the possible time-bin length $\tau$ to fractions of the MZI delay (see Figure \ref{fig:interleaved_TB}). Taking into account the interferometer delay and the known jitter of our detection setup, we choose $\tau = \SI{433}{\pico \second}$. Strong correlations require maximizing the Franson visibility by aligning for high Mach-Zehnder visibilities ($>95\%$) and precisely matching the delays of Alice's and Bob's MZIs. This poses stringent constraints on the alignment: If we assume unit local visibility, the delay mismatch must be below $\approx \SI{10}{\femto \second}$ ($\widehat{=} \ \SI{30}{\micro \meter}$) for a Franson visibility of 99 \%. In practice, we typically observe a Franson visibility of 92 \%, which would correspond to a QBER of roughly 4 \% in the superposition basis.

Behind the analysis stage, photons are coupled into multi-mode fibers and detected by two APDs per module. The detector output pulses are time-stamped by time-tagging modules (TTMs) connected to a GPS-synchronized atomic clock to ensure a common temporal reference frame for both detection modules. Finally, the recorded time tags are sorted into coincidence matrices and further processed.

To maintain stable TSUP correlations despite the sensitivity of Franson interference, we actively phase-lock all MZIs. This stabilization consists of two steps: first, Doppler-free absorption spectroscopy locks the wavelength of the pump laser to a hyperfine structure transition in Potassium \cite{Debs2008} to prevent wavelength fluctuations of the SPDC photons. Second, the pump beam is sent through both MZIs, where classical Mach-Zehnder interference is monitored by two fast photodiodes. Their differential signal feeds back onto the piezo stages where the interferometer mirrors are mounted, forming a loop locking the MZI phases. By manipulating the polarization state of the incoming pump beam, we tune the stabilized phases and select the desired correlations.

\section{III. Methods}

We now turn our attention to the theoretical description of the setup and our methods of analysis. The output state of the source can be written as
\begin{equation}
    \ket{\psi}=\ket{VV} \otimes \int dt f(t)\ket{tt},
\end{equation}
where $f(t)$ is a suitably normalized function determined by the coherence time of the pump laser. Note that polarization DOF is separable. As mentioned above, in order to make practical use of the output state, we choose a suitable discretization of the temporal DOF: While running the experiment, the detection modules record continuous strings of time stamps corresponding to click events at different detectors. Only in post-processing do we introduce the discretization, which casts the underlying continuous state into an entangled qudit state given by
\begin{equation}
\label{state}
     \ket{\psi} = \ket{VV} \otimes \frac{1}{\sqrt{d}}\sum_{i=0}^{d-1} \ket{ii}.
\end{equation}
Realizing the equal superposition of time-bin states requires the pump coherence time to be significantly longer than $d\cdot \tau$. We satisfy this condition by pumping the source with a narrow-band $cw$ laser. This way, we gain access to a quantum state whose dimensionality is, in principle, unbounded. The QKD protocol demonstrated in \cite{bulla2022, bulla2023} pursued a similar strategy, with the crucial difference that they required a hyper-entangled state of the form
\begin{equation}
    \ket{\psi}=\frac{1}{\sqrt{2}}\left(\ket{HH}+e^{-i\phi}\ket{VV}\right) \otimes \frac{1}{\sqrt{d}}\sum_{i=0}^{d-1} \ket{ii}.
\end{equation}
Such protocols demand sources generating hybrid time-bin and polarization entanglement while still achieving high brightness and heralding ratios, which results in increased source complexity and a modest reduction in heralding efficiency. Moreover, accurate alignment of the polarization state after propagating through fibers is tricky, particularly in high-loss scenarios or in varying transmission channels. Compared to this approach, our scheme allows a reduction of the background noise level by $\approx 50\%$ simply by placing a polarizer in front of the measurement apparatus. Now, misalignment of the polarization reference frame between the source and the receivers will only lead to increased transmission losses, while the temporal correlations harnessed for key generation remain unaffected.

After selecting a specific discretization shape, i.e., the length $\tau$ of and distance between time bins and the dimensionality $d$, the first data-processing step is post-selection on meaningful detection events: all time frames that do not contain exactly one detection event at Alice's and one at Bob's side are discarded. The number of invalid click events strongly depends on the chosen discretization. Hence, one must search for a discretization that fits the physical constraints of the setup.

Fortunately, this still leaves a large parameter space spanned by the discretization shape. By exploring this space, we can simultaneously optimize one or several quantities of interest, even if conventionally one expects a trade-off between them. For instance, one may find a sweet spot for the key rate per time and the certifiable entanglement dimensionality. These optimal points depend on the set of environmental conditions given by the photon source, transmission channels, and detection setup. Relevant parameters include source brightness, heralding efficiency, losses, timing resolution, state fidelity, and measurement accuracy. When scanning for the optimal discretization, two limitations must be taken into account:

\textbf{\textit{1.) Parameter Boundary Conditions.}} For instance, if $\tau$ is chosen smaller than the jitter, an increasing number of correlated clicks will fall into different time frames and have to be discarded. Conversely, if $\tau$ or $d$ are chosen too large, more and more uncorrelated click-events will populate identical time frames, forcing additional discarding or leading to errors.

\textbf{\textit{2.) Setup Compatibility.}} The temporal spacing of the time bins must correspond to the physical MZI delay. Otherwise the setup cannot perform TSUP measurements.

These restrictions form the physical background against which our post-processing enables searching for the optimal discretization given dynamically changing conditions or evaluating one dataset for various discretization choices.

\begin{figure*}[htp!]
\begin{center}
\includegraphics[width=1\textwidth]{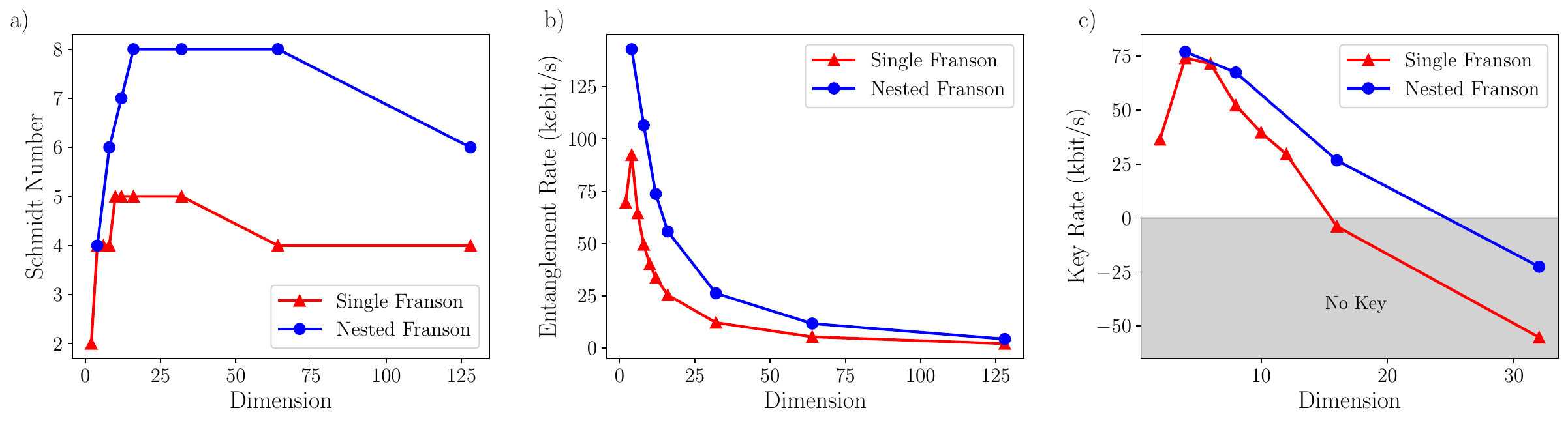}
\end{center}
\caption{Figures of merit for the source and proof-of-principle realization of the novel QKD protocol. Each quantity is computed both for the case when Alice and Bob use only the shorter MZI, effectively a single Franson interferometer, and for the full nested-Franson setup. By varying only the dimension while evaluating the same data set, we scan for optima along one axis of the discretization parameter space. Note that all reported quantities represent lower bounds and are therefore not associated with error bars. a) Schmidt numbers of the photonic state, certifying HD time-bin entanglement for all discretizations with $d>2$. b) Entanglement rate, quantifying the amount of maximally entangled qubits produced per second by the source. c) Asymptotic key rates computed using the novel QKD protocol. Higher dimensions with negative key rates are omitted for clarity. Lines serve as a guide to the eye.}
\label{fig:results}
\end{figure*}

\subsection{A. Entanglement of Formation}
Entanglement of the temporally entangled state can be quantified in terms of the so-called entanglement of formation (EoF), which measures how many $e$bits (i.e., maximally entangled two-qubit pairs) are needed to prepare a copy of the state via LOCC. It can be lower-bounded via
\begin{equation}\label{equ:EoF_bound}
    \text{EoF}(\rho) \geq -\text{log}_2 \left(1-\frac{B(\rho)^2}{2}\right),
\end{equation}
where $B$ is defined as 
\begin{equation}\label{equ:C_in_EoF_bound}
   \begin{split}
    B(\rho) = \frac{2}{\sqrt{|C|}}
    \sum\limits_{\substack{(j,k)\in C \\ j < k}}
    \Big(&
        \left|\bra{jj|\rho}\ket{kk}\right|\\
        &- \sqrt{
            \bra{jk|\rho}\ket{jk}\,
            \bra{kj|\rho}\ket{kj}
        }
    \Big),
\end{split}
\end{equation}
and $|C|$ denotes the cardinality of the set $C$~\cite{Tiranov2017}. $C$ is a subset of the set of all index pairs and can be chosen arbitrarily, which means that it can be optimized such that the lower bound obtains its greatest feasible value.

In practical applications, the number of index pairs that can be considered is limited by the knowledge of the off-diagonal elements of the density matrix. Therefore, a semi-definite programming (SDP), which takes the diagonal and off-diagonal elements determined by the TOA and TSUP measurements as constraints, is employed ~\cite{Martin2017}, which leads to a reliable lower bound on the EoF. We continue to calculate the \emph{entanglement rate} by multiplying the EoF with the coincidence rate in the TOA setting.

\subsection{B. Schmidt Number}
The Schmidt number quantifies entanglement in the sense that Schmidt number $k = 1$ corresponds to separable states, while higher Schmidt numbers indicate the presence of more entanglement. If the density matrix of state $\rho$ is known, the Schmidt number can be bounded using Lemma 1 in \cite{Terhal_2000}.

However, we are interested in high-dimensional states of which we know only the diagonal elements of the density matrix through measurements in the nested-Franson configuration, forcing us to bound the off-diagonals. This would lead to suboptimal Schmidt numbers, as some off-diagonal elements could, in the worst case, be estimated to be negative. To overcome this problem, we introduce a new witness for the Schmidt number, reading
\begin{equation}
    \text{max}_{\rho \in S_k}W(\rho)\leq k
    \label{eq:witness}
\end{equation}
with
\begin{equation}
    W(\rho) = \sum\limits_{i,j=0}^{d-1}\left|\braket{ii|\rho}{jj}\right|,
\end{equation}
which we derive in Appendix A. Similar to the reconstruction of $\rho$ for the entanglement of formation, we solve Eq.~\eqref{eq:witness} with an SDP, minimizing the objective function $W(\rho)$ in the variable $\rho$ subject to constraints deducted from the TOA and TSUP measurements. If this evaluates to $W(\rho)>k$, we have witnessed a Schmidt number of at least $k+1$.

\subsection{C. QKD Protocol and Security}\label{sec:QKD_security}

We implement the following high-dimensional QKD protocol: 

\textbf{\textit{1.) State Generation.}} A photon source produces entangled quantum states $\rho_{AB}$ which are distributed to Alice and Bob.  

\textbf{\textit{2.) Measurement.}} Alice and Bob randomly and independently decide to measure either the time of arrival of the incoming photons (computational basis) or in one of the test bases, where they superpose either the first or the first and second neighboring time bins (see Appendix B). The recorded outcomes are stored in their respective private registers.

Steps 1.) and 2.) are repeated many times. 

\textbf{\textit{3.) Sifting.}} Alice and Bob use the authenticated classical channel to announce their measurement choice. During this step, they may discard certain results or may decide to perform subspace post-selection. 

\textbf{\textit{4.) Statistical Testing.}} In order to estimate correlations between their bit strings, the communicating parties disclose some of their measurement results via the authenticated channel.

\textbf{\textit{5.) Error Correction \& Privacy Amplification.}} Alice and Bob perform error correction and privacy amplification on the remaining rounds such that, in the end, they hold identical keys that are decoupled from Eve.
\\

Certifying the security of high-dimensional QKD protocols is challenging. Most security arguments either rely on mutually unbiased basis measurements~\cite{doda2021, Sheridan_Scarani2010}, which are practically infeasible for many platforms, such as the present one, or on computationally challenging, time-consuming numerical convex optimization tasks that limit the analyzable dimension to the single digits~\cite{bergmayr2023}.

In this work, we use a recent semi-analytic method~\cite{Kanitschar2025, Kanitschar2025b}, which certifies asymptotic secure key rates based on the native measurements of our nested-Franson setup. The underlying idea is the following: while TOA measurements determine the diagonal elements of the unknown state's density matrix, TSUP measurements give access to elements of one or (in the case of our nested setup) multiple off-diagonals.

Inspired by entanglement-witness theory~\cite{Friis2018}, one can construct a witness operator from accessible measurements. Further elements can be bounded using a matrix completion technique~\cite{Tiranov2017, Martin2017}. Eve's guessing probability on Alice's share of the key can be formulated as a convex optimization problem subject to Alice's and Bob's observations on the shared quantum state. A careful combination of the aforementioned techniques can be used to construct an efficiently solvable dual semi-definite program for Eve's guessing probability. The latter, in turn, can be related to the min-entropy of the underlying classical-quantum state,
\begin{equation}
    H_{\mathrm{min}}(X|E)_{\rho} = -\log_2 p_{\mathrm{guess}}(X|E)_{\rho}.
\end{equation}
This leads to a reliable lower bound on the asymptotic Devetak-Winter~\cite{Devetak_Winter2006} key rate,
\begin{equation}
    R^{\infty} \geq H_{\mathrm{min}}(X|E)_{\rho} - H(X|Y),
    \label{eq:keyrate}
\end{equation}
where the classical error-correction leakage $H(X|Y)$ can be bounded based on the test data in the TOA basis. 

\section{IV. Results}
Using the nested-Franson setup, we first probe the entanglement properties of the photons generated by our source. Here, we focus on two operationally meaningful measures: the entanglement dimensionality, or Schmidt number, tests whether we actually succeeded in generating HD entanglement. By computing the entanglement rate, we determine how many maximally entangled qubits, i.e., $e$bits, the source provides per time. By evaluating these quantities from the same dataset for different dimensions $2 \leq d \leq 128$, we scan one axis of the discretization-parameter space. While arbitrary dimensions are in principle possible, we focus on powers of two, as these are the practically relevant dimensions allowing efficient postprocessing.

For the nested-Franson setup, we also superpose the next neighboring time bins, which motivates $4$ as the natural multiple for dimensions. Finally, we calculate lower bounds to the asymptotic key rate given by the novel protocol while again varying the state dimension. Figure \ref{fig:results} summarizes our three figures of merit for the single and nested-Franson setups.

The additional density matrix elements accessible through the nested-Franson setup significantly enhance our capability to witness higher entanglement dimensionalities. This is reflected by the observed entanglement rate, which peaks at $d = 4$, reaching $\approx 92 \ \text{k} e\text{bit}/s$ for the single Franson and $\approx 143 \ \text{k} e\text{bit}/s$ for the nested Franson. The key rate follows a similar trend, with the nested setup providing a smaller but notable advantage of $\approx 74 \ \text{kbit}/s$ versus $\approx 77 \ \text{kbit}/s$ at $d = 4$.

Intuitively, this can be understood as following from the different scaling of the private entropy term and the error correction term of the asymptotic key rate in Eq.~\eqref{eq:keyrate}: While the first contribution scales qualitatively like the entanglement generation and peaks at the dimension where we generate the most entanglement, the second term keeps increasing with growing dimension as errors between Alice and Bob's key strings become more likely.

Crucially, key rate and entanglement rate both exhibit maxima at $d = 4$ already for the single Franson, indicating a sweet spot in the parameter space of possible discretizations where higher dimensions translate into advantages in key rate and entanglement generation, albeit under low loss and noise conditions. \emph{Prima facie}, one would expect strict trade-offs between the certifiable Schmidt number and the entanglement rate: The former typically requires stringent spectral or spatial filtering, which comes at the cost of the latter \cite{Mosley2008, Meyer-Scott2017}. Here, however, we certify Schmidt numbers up to $d = 8$ while simultaneously generating high amounts of entanglement per second. This is possible because our EoF per coincidence, while below the $\log_2(d)$-limit, is offset by the extraordinary brightness of the source.

Whether enhancing the generated entanglement per photon pair, for instance by coupling into single-mode fibers or employing tight spectral filtering, could compensate for the thus reduced coincidence rates remains an open question for future work.

\section{V. Conclusions}
In this work, we present an ultra-bright and stable source of high-dimensional time-bin entangled photons tailored for operation under lossy and noisy conditions. Employing a nested Franson interferometer, we characterize the generated photonic states in terms of entanglement dimensionality, entanglement rate, and achievable key rate within our novel QKD protocol.

Using a flexible evaluation method built around discretizations of the time domain, we, on the one hand, confirm that the additional quantum information accessible through the nested-Franson configuration significantly enhances the certifiable Schmidt number and overall entanglement throughput. On the other hand, we demonstrate optima for our operationally relevant figures of merit beyond the qubit, thus showing a true advantage from going to higher dimensions in key and entanglement rates.

In conclusion, these results highlight the cutting-edge performance of our source design and evaluation techniques, recommending them for demanding entanglement-based quantum communication applications, including, for instance, satellite-based QKD.

\section{Acknowledgements}
F.K., A.B., P.E. and M.H. acknowledge funding from the Horizon-Europe research and innovation programme under grant agreement No. 101070168 (HyperSpace). F.K. gratefully acknowledges support from the Dieberger-Skalicky foundation.

P.E. and M.H. acknowledge funding from the European Research Council (Consolidator grant `Cocoquest’ 101043705) and funding from the Austrian Federal Ministry of Education, Science, and Research via the Austrian Research Promotion Agency (FFG) with the projects FO999914030 (MUSIQ) and FO999921415 (Vanessa-QC) through Quantum Austria.

P.E. and M.H. further acknowledge funding by the European flagship on quantum technologies (`ASPECTS' consortium 101080167). Views and opinions expressed are, however, those of the authors only and do not necessarily reflect those of the European Union. Neither can the European Union be held responsible for them.

\begin{figure}[htb!]
\centering
\includegraphics[width=0.4\columnwidth]{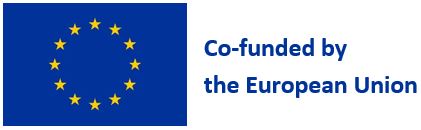}
\end{figure}

\bibliography{bibliography}

\clearpage
\onecolumngrid
\appendix
\section{Appendix A: Proof Of Schmidt-Number Witness}
In this section we prove that for every state $\rho$ on $\mathcal{H} = \mathcal{H}_A \otimes \mathcal{H}_B$ with $\text{dim}(\mathcal{H}_A) = \text{dim}(\mathcal{H}_B) = d \geq 2$ and an orthonormal basis $\{\ket{i}\}_{i=0}^{d-1}$ of $\mathcal{H}_A$ and $\mathcal{H}_B$, we can use
\begin{equation}
    W(\rho):=\sum\limits_{i=0}^{d-1}\sum\limits_{j=0}^{d-1}|\braket{ii|\rho}{jj}|
\end{equation}
as a witness for its Schmidt number, which entails that we have to show
\begin{equation}
    \text{max}_{\rho \in S_k} W(\rho) \leq k,
\end{equation}
where $S_k$ denotes the set of density matrices with Schmidt number at most $k\in \mathbb{N}$. Thus we can conclude that if we observe $W(\rho) > k$, we know that $\rho \notin S_k$, which implies that $\rho$ has Schmidt number at least $k+1$.\\
First, we note that due to convexity we can simplify the maximum over all states with Schmidt number at most $k$ as 
\begin{equation}\label{equ:maxW}
    \text{max}_{\rho \in S_k} W(\rho) = \text{max}_{\ket{\psi_k}}W(\ketbra{\psi_k}),
\end{equation}
where $\ket{\psi_k}$ is a pure bipartite state with Schmidt rank at most $k$, which means it can be written as
\begin{equation}
    \ket{\psi_k} = \sum\limits_{i=0}^{k-1}\ket{i_A i_B},
\end{equation}
where $\{\ket{i_A}\}_i$ and $\{\ket{j_B}\}_j$ are the Schmidt bases of $\ket{\psi_k}$. The equality follows from the Weierstrass Extreme Value Theorem in combination with the Maximum Principle for convex functions, because $S_k$ is a compact, convex set of density matrices with Schmidt number at most $k$ \cite{Terhal_2000} and $W$ is a convex, continuous function of it. We bound the right-hand side of equation \eqref{equ:maxW} as
\begin{equation}
\begin{aligned}
    \text{max}_{\ket{\psi_k}}W(\ketbra{\psi_k}) &= \text{max}_{\lambda}\sum\limits_{i,j=0}^{d-1}\left|\bra{ii}\left(\sum\limits_{l=0}^{k-1}\lambda_l\ket{l_Al_B}\sum\limits_{m=0}^{k-1}\lambda_m\bra{m_Am_B}\right)\ket{jj}\right|\\
    &\leq \text{max}_{\lambda}\sum\limits_{m,n=0}^{k-1}\lambda_m\lambda_n\sum\limits_{i,j=0}^{d-1}\left|\braket{ii}{m_Am_B}\braket{n_An_B}{jj}\right|\\ 
    &= \text{max}_{\lambda}\sum\limits_{m,n=0}^{k-1}\lambda_m\lambda_n\sum\limits_{i=0}^{d-1}\left|\braket{ii}{m_Am_B}\right|\sum\limits_{j=0}^{d-1}\left|\braket{n_An_B}{jj}\right|,
\end{aligned}
\end{equation}
where $\lambda$ denotes the set of possible Schmidt coefficients $\lambda_0, \dotsc, \lambda_{k-1}$ of $\ket{\psi_k}$.

Expressing the bases used in the definition of $W(\rho)$ via the ONBs $\{\ket{x_A}\}_{x=0}^{d-1}$ and $\{\ket{x_B}\}_{x=0}^{d-1}$ to which the Schmidt bases belong, which is possible since the three bases are ONBs of the same Hilbert space, yields
\begin{equation}
    \ket{i} = \sum_x a_x^i \ket{x_A}, \qquad \ket{i} = \sum\limits_x b_x^i \ket{x_B},
\end{equation}
with
\begin{equation}
    \sum_{x}|a_x^i|^2 = 1 \, \forall i, \qquad \sum_{x}|b_x^i|^2 = 1 \, \forall i,
\end{equation}
and
\begin{equation}
     \sum_{i}|a_x^i|^2 = 1 \, \forall x, \qquad \sum_{i}|b_x^i|^2 = 1 \, \forall x,
\end{equation}
because of the orthonormality of $\{\ket{i}\}_i$ and $\{\ket{x_A}\}_x$, $\{\ket{x_B}\}_x$.

We obtain from this via the Cauchy-Schwarz inequality
\begin{equation}
    \sum\limits_{i=0}^{d-1}|\braket{ii}{m_Am_B}| = \sum\limits_{i=0}^{d-1}|a_m^i b_m^i| = \sum\limits_{i=0}^{d-1}|a_m^i| |b_m^i| \leq \sqrt{\sum_i|a_m^i|^2} \sqrt{\sum_i|b_m^i|^2} = 1
\end{equation}
and similarly, 
\begin{equation}
    \sum\limits_{j=0}^{d-1}|\braket{n_An_B}{jj}| \leq 1.
\end{equation}
Lastly, we note that since $\ket{\psi_k}$ is normalized, because Schmidt coefficients are real and non-negative, and again from the Cauchy-Schwarz inequality, it follows that
\begin{equation}
    |\sum\limits_{m=0}^{k-1}\lambda_m \cdot 1|^2 \leq \sum\limits_{m=0}^{k-1}\lambda_m^2 \sum\limits_{m=0}^{k-1} 1^2 = 1 \cdot k  \Leftrightarrow\sum\limits_{m=0}^{k-1}\lambda_m \leq \sqrt{k}.
\end{equation}
Putting things together, we arrive at
\begin{equation}
    \text{max}_{\rho \in S_k} W(\rho) = \text{max}_{\ket{\psi_k}} W(\ket{\psi_k} \leq \text{max}_{\lambda} \sum\limits_{m,n=0}^{k-1}\lambda_m \lambda_n \cdot 1 = \text{max}_{\lambda}(\sum\limits_{m=0}^{k-1}\lambda_m)^2 \leq k,
\end{equation}
which concludes our proof.

\section{Appendix B: Theoretical Analysis of the Nested-Franson Setup}

For a detailed presentation of the theory for the single-Franson measurements and how to reconstruct the density matrix using them, we refer to \cite{bergmayr2023} and focus here on the differences occurring for the nested Franson interferometer, where we have both, measurements after the short interferometer (short-arm) and the long interferometer (long-arm).

Note that the TOA measurements are the same for long-arm and short-arm, so the definition of the ToA clicks stays the same:
\begin{equation}
    \tilde{\textrm{TT}}(i,j) := \textrm{TT}(i,j) = \bra{i,j|\rho_T}\ket{i,j}, \; i,j = 0, 1, \dotsc, d-1
\end{equation}
Same as for the short arm scenario, the TOA clicks ($\tilde{\textrm{TT}}$) already correspond to a POVM,
\begin{equation}\label{eq:P1_POVM1}
  \tilde{\mathcal{M}}_0 := \mathcal{M}_0 =  \left\{ \ketbra{i,j} \right\}_{i,j=0}^{d-1},
\end{equation}
giving rise to the basis $\tilde{\mathcal{B}}_0 := \mathcal{B}_0 = \mathcal{B}_0^{A} \otimes \mathcal{B}_0^B$, where $\tilde{\mathcal{B}}_0^{A/B} := \mathcal{B}_0^{A/B} = \left\{\ket{i}\right\}_{i=0}^{d-1}$ spans single time-bin subspaces. Just like for the short-arm case, the corresponding `basis-click matrix' reads
\begin{equation}
    \tilde{C}_{{\mathcal{M}}_0} := C_{{\mathcal{M}}_0}(i,j) = \textrm{TT}(i,j),
\end{equation}
where we use the natural order $\{\ket{0}, \ket{1}, \hdots, \ket{d-1}\}$.

In contrast, for the TSUP measurements, the definition of the click matrices obviously has to be adapted. We use the notation $\tilde{\textrm{SS}}_{a,b}, \, a,b \in \{1,2\}$ for the TSUP click matrices in the long-arm scenario here:
\begin{align}
    \tilde{\textrm{SS}}_{1,1}(i,j) &= \frac{1}{4} \bra{i\text{++},j\text{++}|\rho_T}\ket{i\text{++},j\text{++}},\, i,j = 2, 3, \dotsc, d-1\\
    \tilde{\textrm{SS}}_{1,2}(i,j) &= \frac{1}{4} \bra{i\text{++},j \mathrm{--}|\rho_T}\ket{i\text{++},j\mathrm{--}},\, i,j = 2, 3, \dotsc, d-1\\
    \tilde{\textrm{SS}}_{2,1}(i,j) &= \frac{1}{4} \bra{i\mathrm{--},j\text{++}|\rho_T}\ket{i\mathrm{--},j\text{++}}, \, i,j = 2, 3, \dotsc, d-1\\
    \tilde{\textrm{SS}}_{2,2}(i,j) &= \frac{1}{4} \bra{i\mathrm{--},j\mathrm{--}|\rho_T}\ket{i\mathrm{--},j\mathrm{--}},\, i,j = 2, 3, \dotsc, d-1,
\end{align}
where $\ket{i/j\text{++}}:=\frac{\ket{i/j} + \ket{i/j - 2}}{\sqrt{2}}$ and $\ket{i/j\mathrm{--}}:=\frac{\ket{i/j} - \ket{i/j - 2}}{\sqrt{2}}$.

To find a corresponding basis $\mathcal{B}_1:=\mathcal{B}_1^A \otimes \mathcal{B}_1^B$, note that $\ket{i\pm\pm}$ for fixed i spans two-dimensional time-
bin subspaces. Restricting us to the case of the dimension $d$ being a multiple of four, we therefore obtain $\{\ket{(4k-2)\pm\pm},\ket{(4k-1)\pm\pm}\}_{k=1}^{\frac{d}{4}} = : \tilde{\mathcal{B}}_1^{A/B}$ for $d=4 l ,\, l \in \mathbb{N}^+$.

Consequently, as one can prove by a short calculation,
\begin{equation}
  \scalebox{0.87}{$
  \begin{aligned}
    \tilde{\mathcal{M}}_1 := &\{\ketbra{(4k-2)\text{++},(4j-2)\text{++}}, \ketbra{(4k-2)\text{++},(4j-2)\mathrm{--}}\}_{k,j = 1}^{\frac{d}{4}}\\
    \cup &\{\ketbra{(4k-2)\mathrm{--},(4j-2)\text{++}}, \ketbra{(4k-2)\mathrm{--},(4j-2)\mathrm{--}}\}_{k,j = 1}^{\frac{d}{4}}\\
    \cup & \{\ketbra{(4k-1)\text{++},(4j-1)\text{++}}, \ketbra{(4k-1)\text{++},(4j-1)\mathrm{--}}\}_{k,j = 1}^{\frac{d}{4}}\\
    \cup &\{\ketbra{(4k-1)\mathrm{--},(4j-1)\text{++}}, \ketbra{(4k-1)\mathrm{--},(4j-1)\mathrm{--}}\}_{k,j = 1}^{\frac{d}{4}}\\
    \cup & \{\ketbra{(4k-2)\text{++},(4j-1)\text{++}}, \ketbra{(4k-2)\text{++},(4j-1)\mathrm{--}}\}_{k,j = 1}^{\frac{d}{4}}\\
    \cup &\{\ketbra{(4k-2)\mathrm{--},(4j-1)\text{++}}, \ketbra{(4k-2)\mathrm{--},(4j-1)\mathrm{--}}\}_{k,j = 1}^{\frac{d}{4}}\\
    \cup & \{\ketbra{(4k-1)\text{++},(4j-2)\text{++}}, \ketbra{(4k-1)\text{++},(4j-2)\mathrm{--}}\}_{k,j = 1}^{\frac{d}{4}}\\
    \cup &\{\ketbra{(4k-1)\mathrm{--},(4j-2)\text{++}}, \ketbra{(4k-1)\mathrm{--},(4j-2)\mathrm{--}}\}_{k,j = 1}^{\frac{d}{4}}
  \end{aligned}
  $}
\end{equation}
forms another POVM, induced by the bases $\tilde{\mathcal{B}}_1^{A/B}$. The corresponding basis-click matrix reads
\begin{equation}
 \scalebox{0.7}{$
\tilde{\mathcal{C}}_{\mathcal{M}_1}:= 
    4\begin{pmatrix}
        \tilde{\textrm{SS}}_{1,1}(2,2) & \tilde{\textrm{SS}}_{1,2}(2,2) & \tilde{\textrm{SS}}_{1,1}(2,3) & \tilde{\textrm{SS}}_{1,2}(2,3) & \tilde{\textrm{SS}}_{1,1}(2,6) & \hdots
         & \tilde{\textrm{SS}}_{1,2}(2,d-2) &
        \tilde{\textrm{SS}}_{1,1}(2,d-1) &
        \tilde{\textrm{SS}}_{1,2}(2,d-1)\\
        \tilde{\textrm{SS}}_{2,1}(2,2) & \tilde{\textrm{SS}}_{2,2}(2,2) & \tilde{\textrm{SS}}_{2,1}(2,3) & \tilde{\textrm{SS}}_{2,2}(2,3) & \tilde{\textrm{SS}}_{2,1}(2,6) &  \hdots  & \tilde{\textrm{SS}}_{2,2}(2,d-2) &
        \tilde{\textrm{SS}}_{2,1}(2,d-1) &
        \tilde{\textrm{SS}}_{2,2}(2,d-1)\\
        \tilde{\textrm{SS}}_{1,1}(3,2) & \tilde{\textrm{SS}}_{1,2}(3,2) & \tilde{\textrm{SS}}_{1,1}(3,3) & \tilde{\textrm{SS}}_{1,2}(3,3) & \tilde{\textrm{SS}}_{1,1}(3,6) &  \hdots  & \tilde{\textrm{SS}}_{1,2}(3,d-2) &
        \tilde{\textrm{SS}}_{1,1}(3,d-1) &
        \tilde{\textrm{SS}}_{1,2}(3,d-1)\\
        \tilde{\textrm{SS}}_{2,1}(3,2) & \tilde{\textrm{SS}}_{2,2}(3,2) & \tilde{\textrm{SS}}_{2,1}(3,3) & \tilde{\textrm{SS}}_{2,2}(3,3) & \tilde{\textrm{SS}}_{2,1}(3,6) &  \hdots  & \tilde{\textrm{SS}}_{2,2}(3,d-2) &
        \tilde{\textrm{SS}}_{2,1}(3,d-1) &
        \tilde{\textrm{SS}}_{2,2}(3,d-1)\\
        \tilde{\textrm{SS}}_{1,1}(6,2) & \tilde{\textrm{SS}}_{1,2}(6,2) & \tilde{\textrm{SS}}_{1,1}(6,3) & \tilde{\textrm{SS}}_{1,2}(6,3) & \tilde{\textrm{SS}}_{1,1}(6,6) &  \hdots  & \tilde{\textrm{SS}}_{1,2}(6,d-2) &
        \tilde{\textrm{SS}}_{1,1}(6,d-1) &
        \tilde{\textrm{SS}}_{1,2}(6,d-1)\\
        \tilde{\textrm{SS}}_{2,1}(6,2) & \tilde{\textrm{SS}}_{2,2}(6,2) & \tilde{\textrm{SS}}_{2,1}(6,3) & \tilde{\textrm{SS}}_{2,2}(6,3) & \tilde{\textrm{SS}}_{2,1}(6,6) &  \hdots  & \tilde{\textrm{SS}}_{2,2}(6,d-2) &
        \tilde{\textrm{SS}}_{2,1}(6,d-1) &
        \tilde{\textrm{SS}}_{2,2}(6,d-1)\\
        \vdots & \vdots & \vdots & \vdots & \vdots & \ddots  & \vdots & \vdots & \vdots \\
        \tilde{\textrm{SS}}_{1,1}(d-2,2) & \tilde{\textrm{SS}}_{1,2}(d-2,2) & \tilde{\textrm{SS}}_{1,1}(d-2,3) & \tilde{\textrm{SS}}_{1,2}(d-2,3) & \tilde{\textrm{SS}}_{1,1}(d-2,6) &  \hdots  & \tilde{\textrm{SS}}_{1,2}(d-2,d-2) &
        \tilde{\textrm{SS}}_{1,1}(d-2,d-1) &
        \tilde{\textrm{SS}}_{1,2}(d-2,d-1)\\
        \tilde{\textrm{SS}}_{2,1}(d-2,2) & \tilde{\textrm{SS}}_{2,2}(d-2,2) & \tilde{\textrm{SS}}_{2,1}(d-2,3) & \tilde{\textrm{SS}}_{2,2}(d-2,3) & \tilde{\textrm{SS}}_{2,1}(d-2,6) &  \hdots  & \tilde{\textrm{SS}}_{2,2}(d-2,d-2) &
        \tilde{\textrm{SS}}_{2,1}(d-2,d-1) &
        \tilde{\textrm{SS}}_{2,2}(d-2,d-1)\\
        \tilde{\textrm{SS}}_{1,1}(d-1,2) & \tilde{\textrm{SS}}_{1,2}(d-1,2) & \tilde{\textrm{SS}}_{1,1}(d-1,3) & \tilde{\textrm{SS}}_{1,2}(d-1,3) & \tilde{\textrm{SS}}_{1,1}(d-1,6) &  \hdots  & \tilde{\textrm{SS}}_{1,2}(d-1,d-2) &
        \tilde{\textrm{SS}}_{1,1}(d-1,d-1) &
        \tilde{\textrm{SS}}_{1,2}(d-1,d-1)\\
        \tilde{\textrm{SS}}_{2,1}(d-1,2) & \tilde{\textrm{SS}}_{2,2}(d-1,2) & \tilde{\textrm{SS}}_{2,1}(d-1,3) & \tilde{\textrm{SS}}_{2,2}(d-1,3) & \tilde{\textrm{SS}}_{2,1}(d-1,6) &  \hdots  & \tilde{\textrm{SS}}_{2,2}(d-1,d-2) &
        \tilde{\textrm{SS}}_{2,1}(d-1,d-1) &
        \tilde{\textrm{SS}}_{2,2}(d-1,d-1)
    \end{pmatrix}.
  $}
\end{equation}

In order to build the last basis, $\tilde{\mathcal{B}}_2^{A/B}$, we have to take into account the remaining TSUP-Klicks ($\tilde{\textrm{SS}}$-Klicks) as well as the clicks from the mismatched measurements ($\tilde{\textrm{TS}}$- and $\tilde{\textrm{ST}}$-Klicks), given by
\begin{align}
    \tilde{\textrm{TS}}_{1}(i,j) &= \frac{1}{2} \bra{i,j\text{++}|\rho_T}\ket{i,j\text{++}}, \\
    \tilde{\textrm{TS}}_{2}(i,j) &= \frac{1}{2} \bra{i,j\mathrm{--}|\rho_T}\ket{i,j\mathrm{--}},\\
    \tilde{\textrm{ST}}_{1}(i,j) &= \frac{1}{2} \bra{i\text{++},j|\rho_T}\ket{i\text{++},j},\\
    \tilde{\textrm{ST}}_{2}(i,j) &= \frac{1}{2} \bra{i\mathrm{--},j|\rho_T}\ket{i\mathrm{--},j}.
\end{align}
There are different possibilities for building a corresponding basis, but if we demand that $d$ is a multiple of four as for $\tilde{\mathcal{B}}_1$, then our third basis is given by $\tilde{\mathcal{B}}_2^{A/B} := \{\ket{0}, \ket{1}, \ket{d-2}, \ket{d-1}\}\cup \{\ket{4k\text{++}},\ket{4k\mathrm{--}}\ket{(4k+1)\text{++}}, \ket{(4k+1)\mathrm{--}}\}_{k=1}^{\frac{d}{4}-1}$.\\

A brief calculation reveals that the POVM induced by $\tilde{\mathcal{B}}_2^{A/B}$ is given by 
\begin{equation}
    {\small
    \begin{aligned}
        \tilde{\mathcal{M}}_2:= &\{\ketbra{i,j}\}_{i,j \in \{0,1,d-2,d-1\}}\\ 
        &\cup \{\ketbra{i,4k\text{\text{++}}}, \ketbra{i,4k\mathrm{--}}\}_{i \in \{0,1,d-2,d-1\}, k=1, \dotsc, \frac{d}{4}-1}\\ 
        &\cup \{\ketbra{i,(4k+1)\text{++}}, \ketbra{i,(4k+1)\mathrm{--}}\}_{i \in \{0,1,d-2,d-1\}, k=1, \dotsc, \frac{d}{4}-1}\\
        & \cup \{\ketbra{4k\text{++},j}, \ketbra{4k\mathrm{--},j}\}_{j \in \{0,1,d-2,d-1\}, k=1, \dotsc, \frac{d}{4}-1}\\ 
        &\cup \{\ketbra{(4k+1)\text{++},j}, \ketbra{(4k+1)\mathrm{--},j}\}_{j \in \{0,1,d-2,d-1\}, k=1, \dotsc, \frac{d}{4}-1}\\
        &\cup \{\ketbra{4k\text{++},4l\text{++}}, \ketbra{4k\text{++},4l\mathrm{--}}\}_{k,l = 1}^{\frac{d}{4}-1}\\
        &\cup \{\ketbra{4k\mathrm{--},4l\text{++}}, \ketbra{4k\mathrm{--},4l\mathrm{--}}\}_{k,l = 1}^{\frac{d}{4}-1}\\
        &\cup \{\ketbra{(4k+1)\text{++},(4l+1)\text{++}}, \ketbra{(4k+1)\text{++},(4l+1)\mathrm{--}}\}_{k,l = 1}^{\frac{d}{4}-1}\\
        &\cup \{\ketbra{(4k+1)\mathrm{--},(4l+1)\text{++}}, \ketbra{(4k+1)\mathrm{--},(4l+1)\mathrm{--}}\}_{k,l = 1}^{\frac{d}{4}-1}\\
        &\cup \{\ketbra{4k\text{++},(4l+1)\text{++}}, \ketbra{4k\text{++},(4l+1)\mathrm{--}}\}_{k,l = 1}^{\frac{d}{4}-1}\\
        &\cup \{\ketbra{4k\mathrm{--},(4l+1)\text{++}}, \ketbra{4k\mathrm{--},(4l+1)\mathrm{--}}\}_{k,l = 1}^{\frac{d}{4}-1}\\
        &\cup \{\ketbra{(4k+1)\text{++},4l\text{++}}, \ketbra{(4k+1)\text{++},4l\mathrm{--}}\}_{k,l = 1}^{\frac{d}{4}-1}\\
        &\cup \{\ketbra{(4k+1)\mathrm{--},4l\text{++}}, \ketbra{(4k+1)\mathrm{--},4l\mathrm{--}}\}_{k,l = 1}^{\frac{d}{4}-1}.
    \end{aligned}
    }
\end{equation}
The corresponding click matrix $\tilde{\mathcal{C}}_{\mathcal{M}_2}$ is omitted here to conserve space. The basis-click matrices are used for normalizing the corresponding entries of $\tilde{\textrm{TT}}, \tilde{\textrm{ST}}_i, \tilde{\textrm{TS}}_i$, $i=1,2$ and $\tilde{\textrm{SS}}_j$, $j=1,2,3,4$ by dividing them by the sum of all the elements of the basis-click matrix they belong to. Similar to the short-arm case, the elements of $\tilde{\textrm{ST}}_i, \tilde{\textrm{TS}}_i$, $i=1,2$ and $\tilde{\textrm{SS}}_j$, $j=1,2,3,4$ that appear in neither $\tilde{\mathcal{C}}_{\mathcal{M}_1}$ nor $\tilde{\mathcal{C}}_{\mathcal{M}_2}$ are not normalized, which, however, is unproblematic since they are not needed for the subsequent calculations, and the elements of $\tilde{\textrm{TT}}$ that appear not only in $\tilde{\mathcal{C}}_{\mathcal{M}_0}$ but also in $\tilde{\mathcal{C}}_{\mathcal{M}_2}$ are still normalized according to $\tilde{\mathcal{C}}_{\mathcal{M}_0}$.

\end{document}